\documentclass[namedreferences]{kluwer}
\usepackage{graphicx}

\begin{document}
\begin{article}

\begin{opening}
\title{Simulations of Galactic Cosmic Rays Impacts onto the Herschel/PACS Photoconductor Arrays with Geant4
Code}

\author{C. \surname{Bongardo}\email{bongardo@pd.astro.it}}
\institute{INAF--Osservatorio Astronomico di Padova, Vicolo dell'Osservatorio 5, I-35122 Padova, Italy}
\author{P. \surname{Andreani}\email{andreani@oats.inaf.it}}
\institute{INAF--Osservatorio Astronomico di Trieste, via Tiepolo 11, I-34136 Trieste, Italy\\
 Max-Planck fur Extraterrestrische Physik, Postfach 1312, D-85741 Garching, Germany }
\author{G. \surname{De Zotti}\email{dezotti@pd.astro.it}}
\institute{INAF--Osservatorio Astronomico di Padova, Vicolo dell'Osservatorio 5, I-35122 Padova, Italy}


\begin{abstract}
We present results of simulations performed with the Geant4 software code of the effects of Galactic Cosmic
Ray impacts on the photoconductor arrays of the PACS instrument. This instrument is part of the
ESA-Herschel payload, which will be launched in late 2007 and will operate at the Lagrangian L2 point of
the Sun-Earth system. Both the Satellite plus the cryostat (the shield) and the detector act as source of
secondary events, affecting the detector performance. Secondary event rates originated within the detector
and from the shield are of comparable intensity. The impacts deposit energy on each photoconductor pixel
but do not affect the behaviour of nearby pixels. These latter are hit with a probability always lower than
7\%.\\
The energy deposited produces a spike which can be hundreds times larger than the noise. We then compare
our simulations with proton irradiation tests carried out for one of the detector modules and follow the
detector behaviour under 'real' conditions.\\
\end{abstract}


\runningtitle{GCRs impact on PACS photoconductor with Geant4}
\runningauthor{Bongardo et al.}
\end{opening}

\section{Introduction}
The European Space Agency Herschel satellite, scheduled for launch on August 2007 with an Ariane-5 rocket,
will operate at the Lagrangian L2 point of the Sun-Earth system (see ESA web page: {\it
www.rssd.esa.int/SA-general/Projects/Herschel}). Herschel is the fourth cornerstone mission of ESA Horizon
2000 program and will perform imaging photometry and spectroscopy in the far infrared and submillimetre
part of the spectrum. The Herschel payload consists of two cameras/medium resolution spectrometers (PACS
and SPIRE) and a very  high resolution heterodyne spectrometer (HIFI).

The PACS (Photo-conductor Array Camera and Spectrometer) instrument will perform efficient imaging and
photometry in three wavelength bands within 60-300$\mu$m and spectroscopy and spectroscopic mapping with
spectral resolution between 1000 and 2000 over the entire wavelength range (60-210$\mu$m) or short
segments.\\
PACS is made of four sets of detectors: two Ge:Ga photoconductor arrays for the spectrometer part and two
Si-bolometer arrays for the photometer part. On each instrument side each detector covers roughly half of
the PACS bandwidth (see Poglitsch et al., 2004). More about Herschel - PACS can be found in the PACS Web
pages: {\it pacs.mpe-garching.mpg.de} and {\it pacs.ster.kuleuven.ac.be}.

It is well known that cosmic rays may influence strongly the detector behaviour in space. The performances
of detectors such as ISOCAM and ISOPHOT on board of the Infrared Space Observatory (ISO) were largely
affected for time periods long enough to corrupt a large amount of data. Our goal, in the present paper, is
to exploit our present knowledge on detectors and cosmic environment to understand how the detector
performances change from the expectations.

In this paper we focus our attention on the GeGa photoconductor arrays (hereafter PhC) of the PACS
instrument, following its response under simulated cosmic rays irradiation. The detectors are similar to
those on board of ISO and we expect therefore that they will be strongly affected by cosmic ray hitting.

In addition we compare proton irradiation tests, performed on a single module of the PhC, with Geant4
simulations and draw conclusions about their performances.

The paper is organized as follows: in Sections \ref{simulations} through \ref{PL}, the used simulation
toolkit, the input detector design, the Galactic Cosmic Ray spectra and the Physics List are outlined. In
Section \ref{results} we report the results. Section \ref{IPN} deals with the irradiation tests which were
performed at UCL-CRC on one detector module and the comparison with our simulations. In the final Section \ref{discussion} we summarize the results reported in this draft.

\section{Geant4 Monte Carlo code}\label{simulations}
Monte Carlo simulations were accomplished with the Geant4\footnote{Every simulation was run with the 7.0
patch 01 version, with CLHEP 1.8.1.0 with a gcc 3.2.3 compiler} (hereafter G4) toolkit, which simulates the
passage of particles through matter. G4 provides a complete set of tools for all the domains of detector
simulation: Geometry, Tracking, Detector Response, Run, Event and Track management, Visualization and User
Interface. An abundant set of Physics Processes handle the diverse interactions of particles with matter
across a wide energy range, as required by the G4 multi-disciplinary nature; for many physical processes a
choice of different models is available. For any further information see the Geant4 Home page at: {\it
geant4.web.cern.ch/geant4/}.

\section{The Detector Design}\label{volumes}
\subsection{PACS as a spectrometer}
The PACS employs two PhC arrays (stressed/unstressed) to perform imaging line spectroscopy and imaging
photometry  in the 60 - 300 $\mu$m wavelength band. In spectroscopy mode, it will image a field of about 50
x 50 arcsec, resolved into 5 x 5 pixels, with an instantaneous spectral coverage of $\approx$1500 km/s and
a spectral  resolution of $\approx$175 km/s with an expected sensitivity (5$\sigma$ in 1h) of 3 mJy or
2.5$\times 10^{-18}$ W/m$^2$.

The high stressed module is cooled down to 1.8 K (whereas the low stressed is cooled to 2.5 K), the
Front End Electronic (FEE) to 4 K. All main metal components are made of a light metal alloy called
Erg-Al (AlZn5,5MgCu).

\subsection{The Geometry}\label{Geometry}
\begin{figure*}
\centerline{\includegraphics[height=8cm, width=6cm]{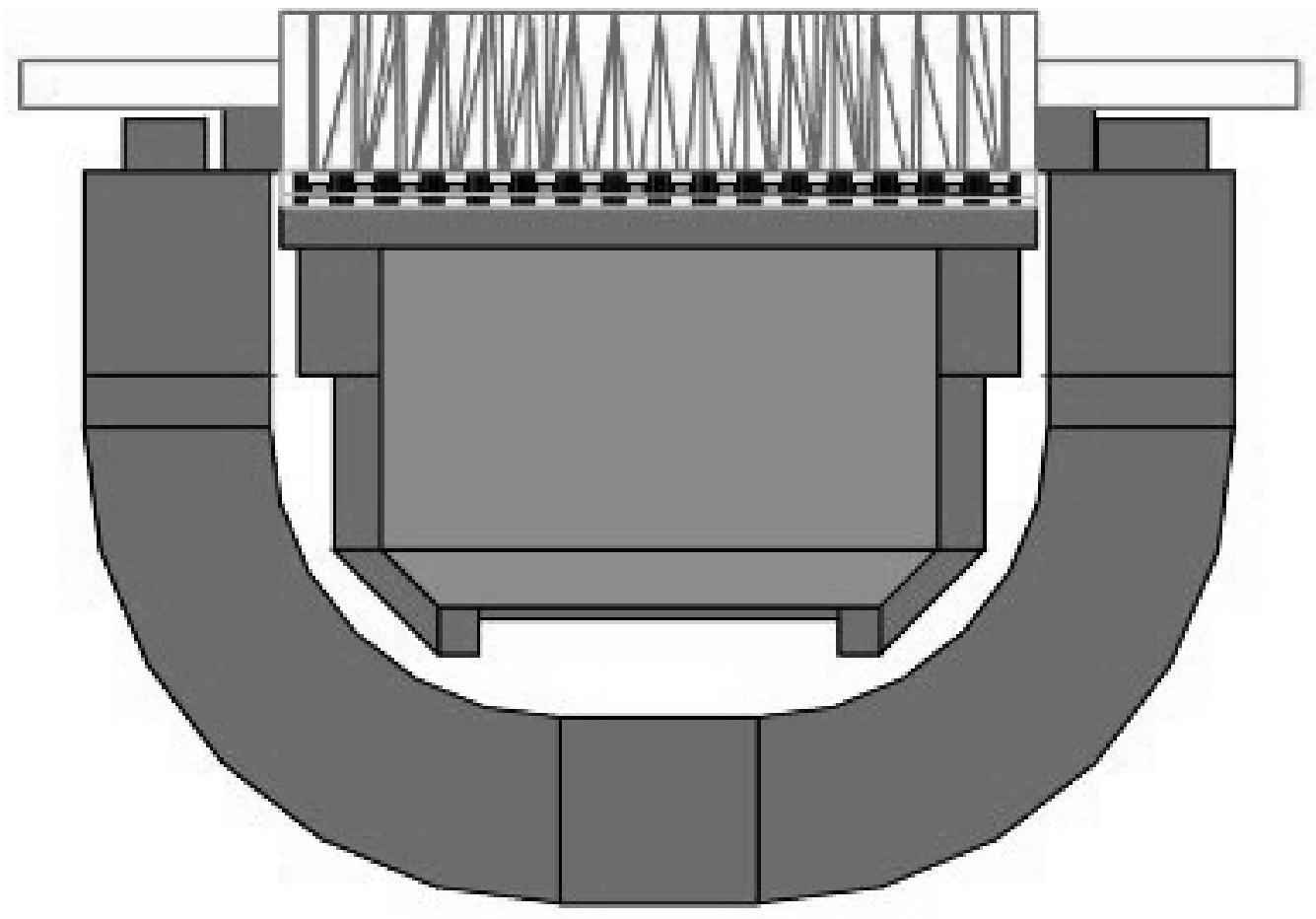}
\includegraphics[height=8cm, width=6cm]{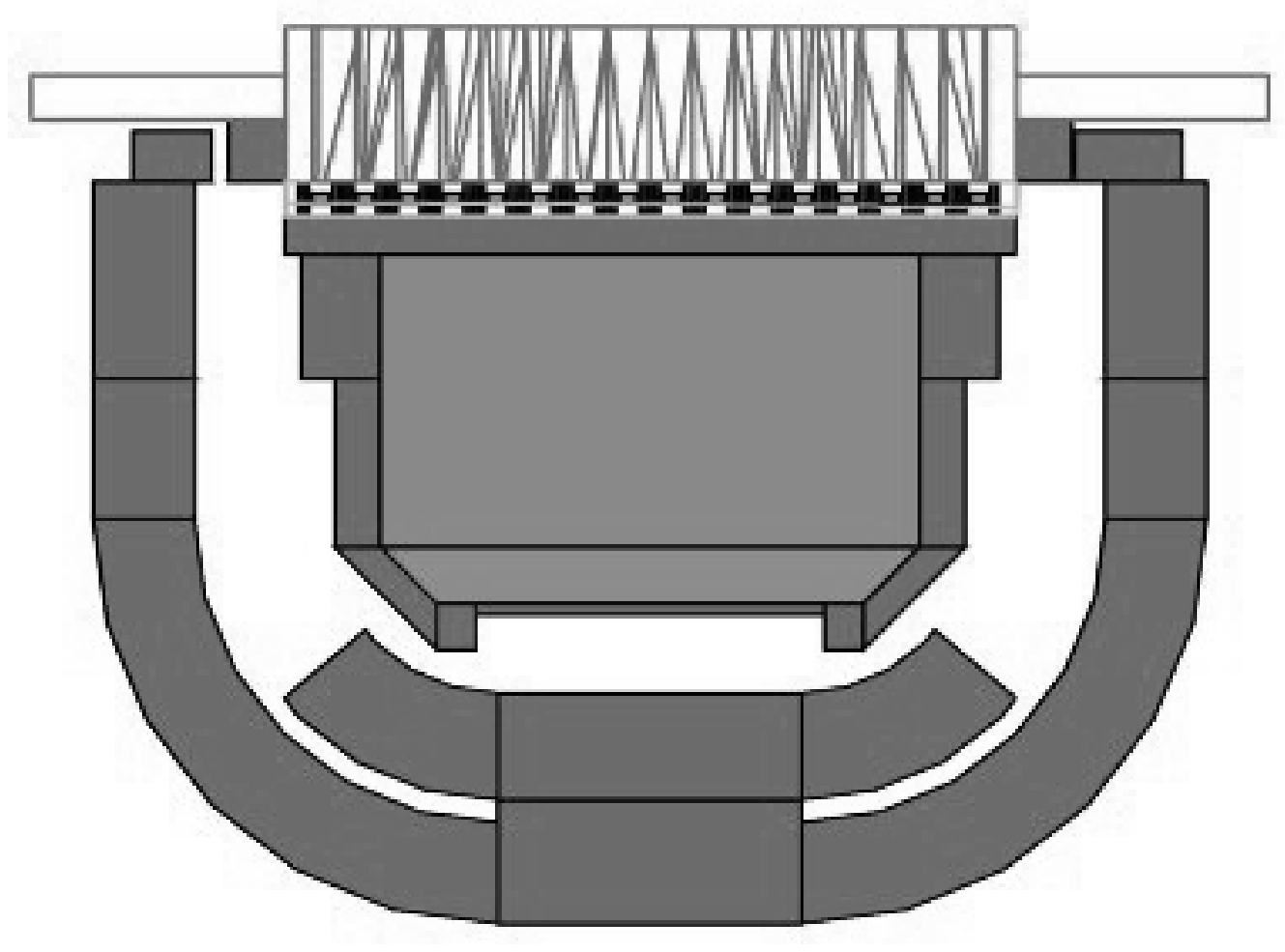}}
\caption[]{The PACS high (left) and low (right) stressed module (left).\\
Magenta: the fore-optics. Yellow: the gold coating of the FEE and the active pixel. Red:  the CuBe2
contacts. Blue: the Al$_2$O$_3$ insulator. Gray: the cylindrical steel segment. Cyan: Erg-Al
components.}\label{high}
\end{figure*}

\begin{figure*}
\centerline{\includegraphics[height=10cm, width=8cm]{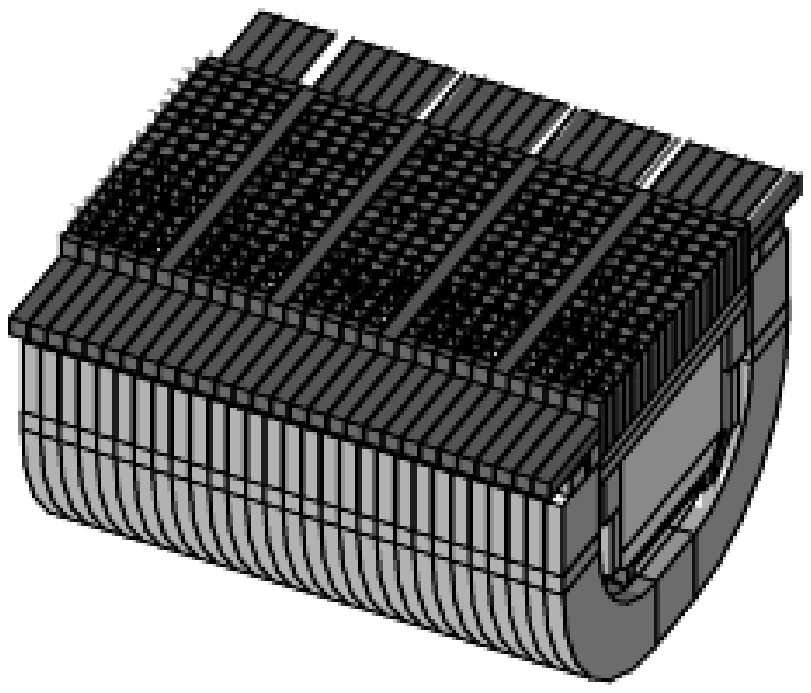} }
\caption[]{The PACS array. It contains 25 modules of 16 pixels each, making an array of 25x16.}\label{PhC}
\end{figure*}

Figures \ref{high} and \ref{PhC} show the modeled detectors with Geant4.\\
The most important part of the PhC is the cavity. Inside the cavity there is the active part of the PhC,
the pixel. Each pixel is made of Ge:Ga (in the high stress configuration Ga doping is 10$^{14}$ cm$^{-3}$,
that is an atom of Ga for about 10$^9$ atoms of Ge; this negligible Ga concentration is not further
considered here). The pixel is a simple box of 1$\times$1$\times$1.5  mm. Each pixel is sustained by two
boxes (`cubes') made of CuBe2 (a Cu, Be and Pb alloy). On  the external parts of these two last boxes there
are two different cylinders made of CuBe2 (pedestal 1 and 2). A steel cylinder (the {\it kugel}) is placed
on the right-hand side of pedestal 1. A thinner and smaller cylinder is located in the left-hand side. The
sequence of a single pixel is: {\it kugel}, pedestal 1, cube, pixel, cube and pedestal 2 with its thinner
cylinder. Between two sequences of pixels there is an Al$_2$O$_3$ (Aluminum oxide) insulator. The sequence
of the 16 pixels starts with an insulator and ends with a {\it kugel}. The pixels have a single face that
is not shadowed by the cavity, that is, clearly, the one facing the fore optics (see below).

The fore optics (hereafter FO) is the most difficult part to be designed with G4. Actually it is
constructed as a unique block made of Erg-Al, from which they subtracted a cone non orthogonal to the
respective pixel. We designed each cone individually, making use of the boolean solids. All the FO should
be gold coated with a 8.95 g gold mass. Simple calculations of the surface area of the FO leads to a
coating thickness of $\approx$1.6 $\mu$m. We did not reproduce such a coating.

The FEE is made of two different volumes: a box and a trapezoid, both 0.5 mm thick. The Erg-Al components
of the module have three different thicknesses. The small cubes sustaining the FO are 1.75 mm thick; the
end of the horseshoe is 2.3 mm thick; the  horseshoe is 4.1 mm thick.

Figure \ref{high} shows the geometry of the PACS PhC (the single stressed and unstressed modules), as drawn
in G4.  Figure \ref{PhC} shows the whole stressed PhC detector. This geometry is that used by the G4
software for the simulations.

For the Galactic Cosmic Rays simulations we considered all the components `external' to the detector, i.e.
the PACS box, the cryostat and the spacecraft and called the ensemble `shield' and approximated it with an
Aluminum sphere, with thickness of 11 mm and at a temperature of 80 K, which is the nominal temperature of
the passive cooled Herschel telescope. The inner radius is 20 cm and the outer 21.1 cm.

For the irradiation tests the external environment was modelled as close as possible to the real one and it
is described in details in \S\ref{IPN}.

The G4 software then requires the design of a `world volume' containing all the detector part and the
specification of the environment in which the detector is placed. We made the world volume a little bit
larger than the shield, i.e. is a sphere with a radius 21.2 cm.

\section{Galactic Cosmic Rays spectra}
Herschel will be injected in a orbit around L2. The L2 environment, as regards the Galactic Cosmic Ray
(hereafter GCR) rates, has been measured by the WMAP satellite \cite{Prantzos}. This environment is similar
to that of the geostationary orbit and therefore the GCR spectra are known. In the limit of the distance
Earth-L2 the GCRs flux gradient is negligible. Therefore we use GCR fluxes of a geostationary orbit as a
reasonable approximation. In particular we use the Cosmic Ray Effects on Micro-Electronics (CREME) model
set both for the solar-minimum (GCR maximum) and for the solar-maximum environment \cite{CREME}. This model
uses numerical models of the ionizing radiation environment in near-Earth orbits to evaluate the resulting
radiation effects on electronic systems in space\footnote{The CREME version used in this work is CREME94,
which was more extensively tested and compared with measurements, however there is an updated version,
CREME96, with a web page: {\it https://creme96.nrl.navy.mil/}}.

A very diffuse background, called {\it galactic vacuum} by G4, which is a statistical view of the real
interactions and diffusions of the Galactic Cosmic rays sources with and from the interstellar material,
is defined and used as a good approximation of an empty space where an isotropic cosmic ray flux is
present. This environment has the following physical properties:

\begin{description}
\item A = 1.01 g/mole (atomic mass)
\item Z = 1 (atomic number)
\item $\rho$ = 1$\times$10$^{-25}$ g/cm$^3$ (density)
\item P = 1$\times$10$^{-19}$ Pa (pressure)
\end{description}

\begin{figure*}
\centerline{
\includegraphics[height=6cm, width=6cm]{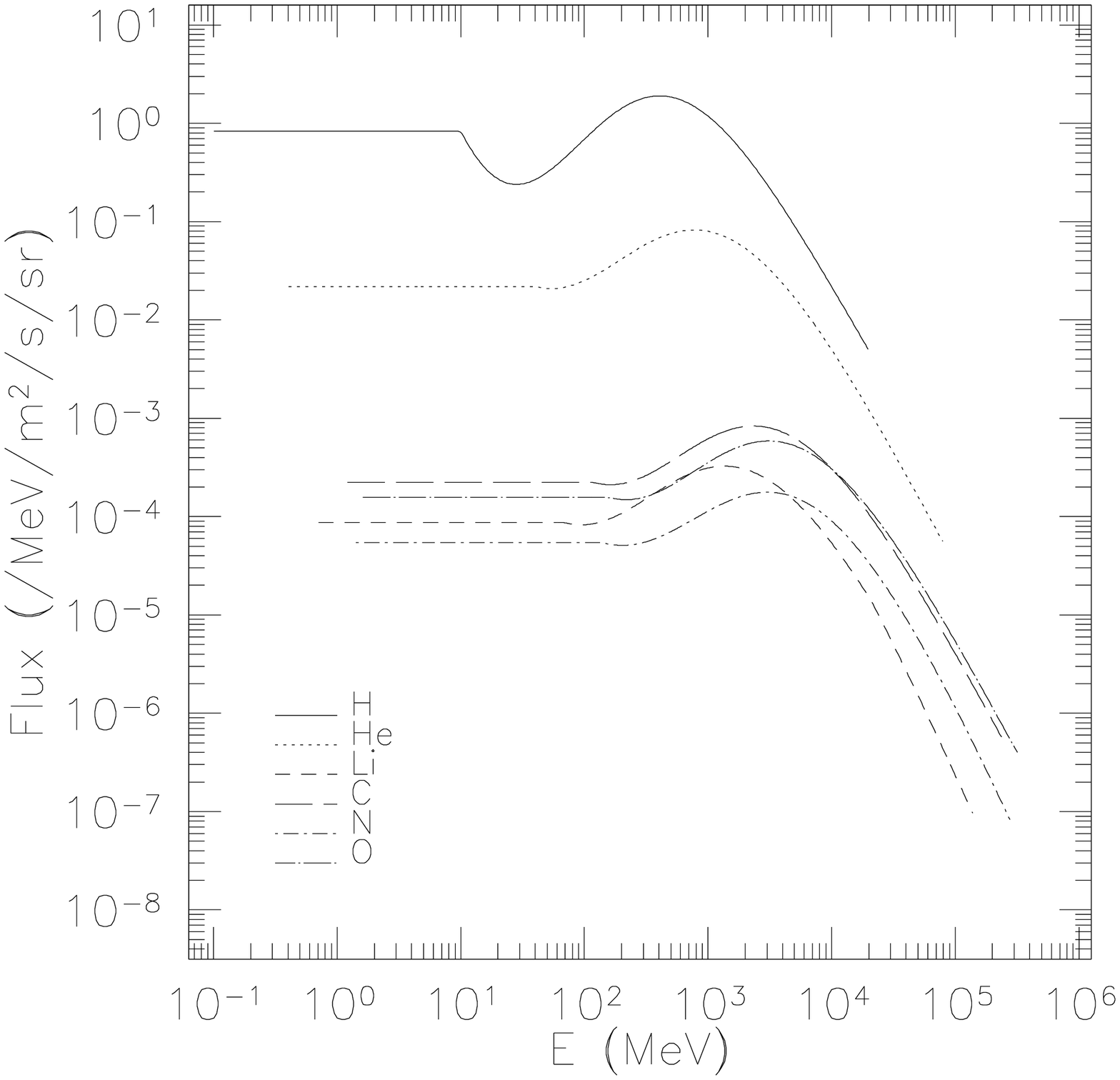}\\
\includegraphics[height=6cm, width=6cm]{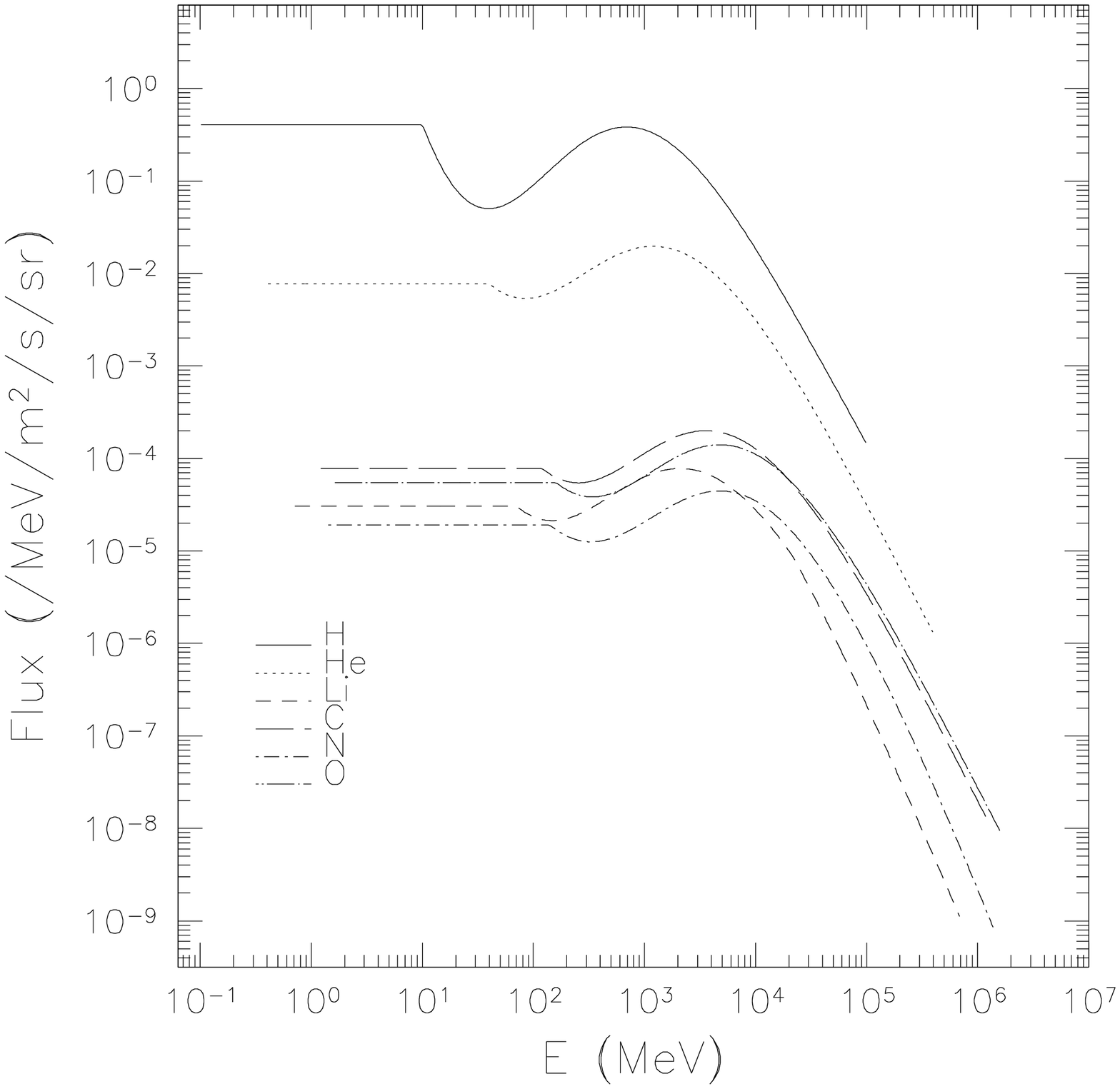}}
\caption[]{\small The differential GCR SEDs at the solar minimum (left) and maximum (right). Each curve
corresponds to a type of particle (from top to bottom: proton, $\alpha$, C, O, Li, N). The rates reported
in Table \ref{GCR_rates} are the integrals over energy and refer to the full shield
surface.}\label{GCR_SED}
\end{figure*}

Our simulation considers the most common GCRs in the near orbit Earth heliosphere: protons, alpha particles
and nuclei of Li, C, N, O (see Figure \ref{GCR_SED}). It cannot include heavier nuclei, such as iron, since
Fe physics has not yet been implemented in the Ion Physics of G4 (see \S\ref{PL}). In Table \ref{GCR_rates}
we report the rates (number of particles per second) of each GCR type computed by integrating the fluxes
reported in figure \ref{GCR_SED} over the shield surface and particle energy. Herschel will reach the L2
point in 2007 at a solar minimum but it will operate until an epoch of increasing solar activity.
Therefore, we consider both solar activity phases.

\begin{table}
\begin{tabular}{lrr}
\hline
	& Solar min & Solar max \\
GCR Type& rates (\#/s)  & rates (\#/s)  \\
\hline
H      &11275.89    	&4553.28        \\
He     & 1102.56        & 458.67        \\
Li     &    7.46        &   2.96        \\
C      &   33.52        &  13.92        \\
N      &    8.96        &   3.68        \\
O      &   31.31        &  13.00        \\
\hline
\end{tabular}
\caption[]{Rates (number of nuclei per second, on a sphere of 21.2 cm radius) for different GCR types
and solar activity.}\label{GCR_rates}
\end{table}

\subsection{The Generate Source Particle Module}
G4 is built in such a way that once the geometry (volumes) and the physics are chosen and setup, the
tool may be played from outside the main program through an User Interface (UI) easy accessible and
modifiable (i.e. a macro file). In this macro file the user specifies: (a) the graphical
visualization and/or (b) the type and the number of particle hitting the detector and their energy. One
single particle with a determined energy represents an event.

We simulated the GCR seedings via the General Source Particle Module (GSPM; Ferguson 2000a,b,c), a code for
the spacecraft radiation shielding, that is based on the radiation transport code G4. The user must specify
the input parameters of the particle source: (1) the spatial and (2) the energy distribution. Every time G4
requires a new event, the GSPM randomly generates it according to the specified distributions. We adopt an
angular distribution that is isotropic over a spherical surface (the vacuum sphere), without any preferred
particle momentum. We then set the energy distribution of each particle according to the SEDs shown in
Figure \ref{GCR_SED} (see also \S\ref{theorcomp}).

\section{The Physics Model}\label{PL}
We had to define a list of physical interactions with the shielding and the different components of the
photoconductors in order to best reproduce the impact of GCRs on the photoconductors. The list is
comprehensive of all the possible physical interactions:

\begin{itemize}
\item Electro-magnetic physics: photo electric and Compton effect; pair production; electron and
positron multiple scattering, ionization, bremsstrahlung and synchrotron; positron annihilation.
\item General physics: decay processes.
\item Hadron physics: neutron and proton elastic, fission, capture and inelastic processes; photon,
electron and positron nuclear processes; $\pi^+$, $\pi^-$, K$^+$, K$^-$, proton, anti-proton,
$\sigma^+$, $\sigma^-$, anti-$\sigma^+$, anti-$\sigma^-$, $\chi^+$, $\chi^-$, anti-$\chi^+$,
anti-$\chi^-$, $\omega^+$, $\omega^-$, anti-$\omega^+$ and anti-$\omega^-$ multiple scattering and
ionization. We added the High Precision Neutron dataset, that is valid for neutrons till 19.9 MeV.
\item Ion physics: multiple scattering, elastic process, ionization and low energy inelastic processes.
We added inelastic scattering (J.P. Wellish, private communication) using the binary light ion reaction
model and the Shen cross section \cite{Shen}.
\item Muon physics: $\mu^+$, $\mu^-$ multiple scattering, ionization, bremsstrahlung and pair
production; $\mu^-$ capture at  rest; $\tau^+$ and $\tau^-$ multiple scattering and ionization.
\end{itemize}

For 'ion physics' of $alpha$-particles, protons, and H isotope nuclei we used the cross section by Tripathi
et al. (2000), holding for energies of up to 20 GeV per nucleon. The cross sections used for the other ions
\cite{Shen} are valid only up to 10 GeV per nucleon. This means that we had to cut out high energy ions of
the GCRs (whose flux is lower than 10$^{-5}\,$/m$^2$/s/sr). The treatment of Ion Physics is strictly valid
only for nuclei not heavier than C. However we have applied it also to simulate N and O, which should not
behave much differently.  

The model called by our physics list is different depending on the energy of the incoming particles. Our
physics list is based on QGSP\_BIC\footnote{Quark Gluon String Physics\_Binary Cascade} list (J.P. Wellish,
private communication; see the Geant4 web page for details). Modifications to such a list are:

\begin{enumerate}
\item We added the High Precision Neutron dataset, which is valid for neutrons up to 19.9 MeV.
\item For neutrons and protons, we introduced\footnote{It is well known that (1) the BIC
does not reproduce well data for energies below 50 MeV \cite{Ivan03}. In contrast, the Bertini
cascade works well below 50 MeV for all but the lightest nuclei. And (2) the BIC for pions and kaons
has not been tested yet (J.P. Wellish, private communication).} the Bertini cascade \cite{Bertini} up to 90
MeV and made the BIC starting from 80 MeV.
\item QGSP is assumed valid from 10, rather than 12, GeV.
\item We have taken into account the gamma- and electro-nuclear reactions, so we considered the
electromagnetic\_GN physics list.
\end{enumerate}


\subsection{Thresholds}
In order to avoid the infrared divergence, some electromagnetic processes require a threshold below which
no secondary particles will be generated. Then, each particle has an associated production threshold, i.e.
either a distance, or a range cut-off, which is converted to an energy for each material, which should be
defined as an external requirement. A process can produce the secondaries down to the recommended
threshold, and by interrogating the geometry, or by realizing when the mass-to-energy conversion can occur,
recognize when particles below the threshold have to be produced.

We have set different production thresholds (hereafter cuts) for each geometrical region.\\
We decided: 1) not to use very small cut values (below 1-2 $\mu$m), since this could affect the validity
of physical models at such small steps (V. Ivantchenko, private communication); 2) to set cut values
$\approx$ 1/5 of the volume thickness. We identified groups of volumes with the same thicknesses: in
particular we defined 6 different regions (see Table \ref{cuts}).\\
Results are robust since they do not depend on the cut values, both default and the chosen ones lead to
similar outputs, with an increasing of the secondary event production.  The results (energy deposition)
among the different simulations were always comparable at 1$\sigma$. On the one hand this was expected,
since most of the volumes have still the same default cut value, 0.7 mm. On the other hand it is possible
that Erg-Al thresholds have already achieved convergence at the default values, that is we are very
slightly tuning them. Although results are pretty similar (see Table \ref{nospPD}) those with our chosen
cuts allow a better tracking inside each detector volume.

\begin{table}
\begin{tabular}{lrl}
\hline
Region      & Cut Value     & Thickness \\
\hline
Shield          & 2 mm          & 11 mm         \\
High\_mm        & 0.7 mm        & 3.02 - 4.1 mm \\
Low\_mm         & 0.3 mm        & 0.73 - 2.1 mm \\
High\_mic       & 100 $\mu$m    & 300 - 600 $\mu$m\\
Low\_mic\_a     &   5 $\mu$m    & 20 $\mu$m     \\
Low\_mic\_b     &   3 $\mu$m    & 10 $\mu$m     \\
\hline
\end{tabular}
\caption[]{\small The five regions of the photoconductor array and their specific cut
values.}\label{cuts}
\end{table}

\subsection{Theoretical computation of GCR flux}\label{theorcomp}
In order to disentangle the contributions from the different simulated parts of the instrument (detector
and shield) we evaluated the impact of the GCR flux onto the `naked' detector (without spacecraft
shielding). In Table \ref{nospPD} we report the results from the simulated GCR flux hitting the detector
only.\\
We computed the number of impacts, $n_{\rm impacts}$, as $n_{\rm impacts} = \pi\Phi S$, where $\Phi$ is the
incoming particle flux in cm$^{-2}$ s$^{-1}$ sr$^{-1}$), and $S$ is the total surface area of the volume
considered.

The hit average values in the two cases (with G4 default cut values and with our cuts) are compared to
those compute theoretically.\\
Fixing specific cuts increases the error bars, but $\langle n_{\rm impacts}\rangle$ comes closer to the
expected mean value. The chosen cut values are a fine tuning of the Erg-Al cuts and, as a result, this
leads to a very precise tracking inside the detector volumes.

\begin{table}
\begin{tabular}{lccc}
\hline
	&	& Default Cuts  & Chosen Cuts   \\
Volume  & Th. Hit  & $\langle n_{\rm impacts}\rangle$  & $\langle n_{\rm impacts}\rangle$ \\
\hline
Pixel   &56.43  &50.45$\pm$10.34        &58.18$\pm$12.81    \\
\hline
\end{tabular}
\caption[]{\small Comparison between the average values of the hit numbers in three distinct cases:
theoretical computation (see \S \ref{theorcomp}), G4 simulation without cuts, G4 simulation with cuts.
This computation was done for the high stressed PhC without shield and secondary events and for the
pixel volume.}\label{nospPD}
\end{table}

\section{Results from GCR simulations}\label{results}

\subsection{Preliminary Tests}
To test the correctness of our results we first made some preliminary tests.

\begin{figure}
\centerline{\includegraphics[height=8cm, width=8cm]{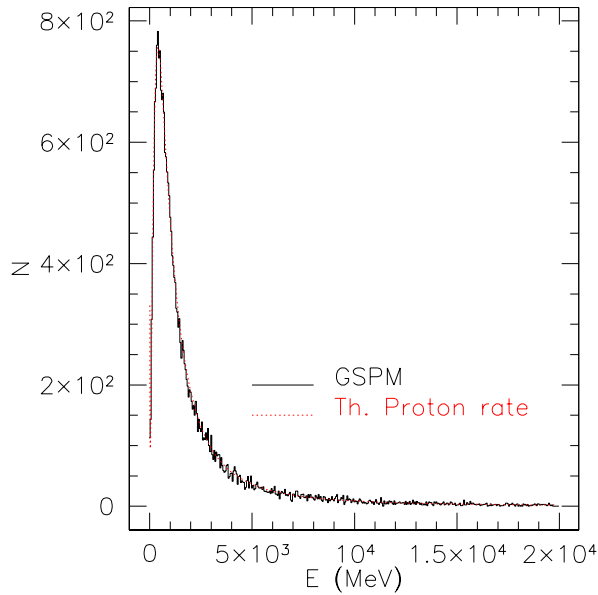}}
\caption[]{\small The superposition of theoretical proton rate at solar minimum (dotted line) with the
sum of five simulations with GSPM (continuous line). GSPM reproduces very well the GCR spectral energy
distribution.}\label{enGSPM}
\end{figure}

The goodness of the internal randomness of GSPM was checked as follows. In Figure \ref{enGSPM} both the
theoretical (CREME94) and the G4 output of the GSPM energy distributions are shown. Five simulations
with 5$\times$10$^3$ particles each were run and the composite output is compared to the theoretical
normalized one (see Figure \ref{GCR_SED}).

\subsection{The $n_{\rm sec}/n_{\rm prim}$ ratio}
Table \ref{HLnuclei} reports the number of secondary particles produced on average per primary event. The
$n_{\rm sec}/n_{\rm prim}$ ratio is clearly increasing with the atomic number of the primary particle and
the high and low stressed PhC have rather the same dependence. Unfortunately such a ratio is strictly
dependent on the chosen physics list and on the reliability of the G4 models. G4 has been tested against
nuclear data for nuclei lighter than or equal to C, therefore the ratios results on N and O must be taken
with caution.

\begin{table}
\begin{tabular}{lcccc}
\hline
    	& \multicolumn{2}{c}{High Stressed} & \multicolumn{2}{c}{Low Stressed}  \\
    	& Solar min     & Solar max & Solar min & Solar max     \\
GCR     & $\langle n_{\rm sec}\rangle/n_{\rm prim}$ & $\langle n_{\rm sec}\rangle/n_{\rm prim}$
    	& $\langle n_{\rm sec}\rangle/n_{\rm prim}$ & $\langle n_{\rm sec}\rangle/n_{\rm prim}$     \\
\hline
H       & 3.302$\pm$0.122       & 4.488$\pm$0.157
        & 3.272$\pm$0.088       & 4.487$\pm$0.127\\
He      & 6.569$\pm$0.265       &10.612$\pm$0.279
        & 8.672$\pm$0.241       &10.385$\pm$0.335\\
Li      &14.665$\pm$0.744       &17.409$\pm$0.416
    	&14.833$\pm$0.269   	&17.394$\pm$0.493\\
C       &35.859$\pm$0.818       &45.489$\pm$0.766
    	&36.473$\pm$0.913   	&45.844$\pm$1.105\\
N       &45.527$\pm$1.819       &57.420$\pm$0.913
    	&46.275$\pm$1.212   	&57.658$\pm$1.015\\
O       &55.863$\pm$1.394       &70.275$\pm$1.188
    	&55.379$\pm$1.745   	&72.563$\pm$1.234\\
\hline
\end{tabular}
\caption[]{\small The $n_{\rm sec}/n_{\rm prim}$ ratio from the simulation of GCRs radiation on the
photoconductors. In the first column the GCR type; in column \#2 and \#3 the average of secondary events
generated by a single primary event for the high stressed PhC in case of solar min and max irradiation; in
column \#4 and \#5 same as \#2 and \#3, but for the low stressed PhC.}\label{HLnuclei}
\end{table}

\subsection{Deposited energy, definition of a glitch}
A glitch on the photoconductor is an unexpected voltage jump during the integration ramp. Along a single
ramp, that lasts 300 ms, the voltage with a 64 readout sequences decreases monotonically. When the voltage
jump exceeds by 4 -- 6 times (in sigma units) the mean jump we have a glitch.

The exact electronic behaviour is not reproducible with Geant4, we only collected the energy depositions on
the pixels. The energy deposited on the pixels is transformed \cite{MG} in detector signal (Volt):

\begin{equation}\label{EdV}
\Delta E(MeV)= C\Delta V(Volt)E_g/(Rh\nu/\eta)
\end{equation}

\noindent
where the photoconductive gain R = 30 A/W, at $\lambda$=170 $\mu$m ($\eta$ = 0.3, with a large error
bar); E$_g$ is the energy gap, and it is equal to 2.9 eV, and C is the detector capacity. For C = 3 pF,
we get: $\Delta V = 0.0134 \Delta E$.

Impacts on pixels are identified and used to compute the deposited energy. Tables\footnote{Tables report
the sum of the voltage jumps normalized to a second of GCR impacting. This value must be corrected
by the readout and ramp lasting.} \ref{HdV} and \ref{LdV} report the mean deposited energy (in V) for
both (high and low) stressed PhC for shielded simulations, i.e. those simulating the effect of the
spacecraft. 

\subsubsection{High stressed Photoconductor}
We considered first the effects on the high stressed PhC of the primary particle impacts on the
shield and the detector, and the production of secondary events from the shield and the detector. We ran
five simulations for each nucleus. Each simulation considers 2$\times$10$^5$ particles. Results (see
Table \ref{HdV}) have been normalized to 1 s of irradiation.

\begin{table} 
\begin{tabular*}{\maxfloatwidth}{lcccc} 
\hline
                & \multicolumn{4}{c}{High Stressed pixel}         \\
GCR             & \multicolumn{2}{c}{Solar min}         & \multicolumn{2}{c}{Solar max} \\
                & $\langle n\rangle$                    & $\langle dE_{dep}\rangle (MeV)$ 
                & $\langle n\rangle$                    & $\langle dE_{dep}\rangle (MeV)$ \\ 
\hline 
H               &101.41$\pm$5.68      			&60.21$\pm$3.19
                & 50.86$\pm$0.84			&27.35$\pm$1.83                \\

He              & 17.46$\pm$0.66      			&23.72$\pm$0.88
                &  8.77$\pm$0.23			& 9.31$\pm$0.09                \\
      
Li              & 0.019$\pm$0.001      & 0.32$\pm$0.01
                & 0.089$\pm$0.004      & 0.12$\pm$0.01	\\
 
C               &  2.24$\pm$0.07      			& 4.64$\pm$0.18 
                &  1.15$\pm$0.04			& 1.86$\pm$0.09                \\

N               &  0.76$\pm$0.02      & 1.54$\pm$0.06 
                &  0.39$\pm$0.01      & 0.64$\pm$0.01	\\

O               & 3.23$\pm$0.09			   	& 7.07$\pm$0.27
                & 1.74$\pm$0.04				& 2.89$\pm$0.15	\\
\hline 
\end{tabular*} 
\caption[]{\small Mean number of hits and energy deposition expressed in MeV, for the high stressed pixels
with spacecraft shielding, secondary events at solar minimum and maximum.}\label{HdV}
\end{table}

\subsubsection{Low Stressed Photoconductor}
We ran five simulations for each nucleus. Each simulation considers 2$\times$10$^5$ particles. Results
(see Table \ref{LdV}) have been normalized to 1 s of irradiation.

\begin{table} 
\begin{tabular*}{\maxfloatwidth}{lcccc} 
\hline
                & \multicolumn{4}{c}{Low Stressed pixel}         \\
GCR             & \multicolumn{2}{c}{Solar min}         & \multicolumn{2}{c}{Solar max} \\
                & $\langle n\rangle$                    & $\langle dE_{dep}\rangle (MeV)$ 
                & $\langle n\rangle$                    & $\langle dE_{dep}\rangle (MeV)$ \\ 
\hline 
H               &105.29$\pm$5.63      			&61.74$\pm$4.42
                & 47.98$\pm$0.95			&24.80$\pm$1.54                \\

He              & 17.59$\pm$0.49      			&24.21$\pm$0.99
                &  8.66$\pm$0.37			& 9.41$\pm$0.59                \\
      
Li              & 0.019$\pm$0.003      & 0.332$\pm$0.001
                & 0.087$\pm$0.001      & 0.119$\pm$0.003	\\
 
C               &  2.22$\pm$0.05      			& 4.62$\pm$0.05 
                &  1.18$\pm$0.03			& 1.90$\pm$0.08      \\

N               & 0.78$\pm$0.01      & 1.63$\pm$0.09 
                & 0.399$\pm$0.001      & 0.64$\pm$0.03	\\

O               & 3.33$\pm$0.05			   	& 7.36$\pm$0.21
                & 1.72$\pm$0.04				& 2.85$\pm$0.19	\\
\hline 
\end{tabular*} 
\caption[]{\small Same as Table \ref{HdV} for the low stressed pixels.}\label{LdV}
\end{table}

\subsection{Cross talks}\label{cross}
We checked whether in the G4 outputs there are coincidences ({\it cross talks}) between adjacent pixels.
The number of pixels affected by one single GCR as resulting from G4 simulations are listed in Table
\ref{Hng} for the high stressed PhC\footnote{No significant dependence on solar activity or on stress
status is found.}. Values are given in percentage for each GCR type producing cross talks. The percent
probability that nearby pixels are affected by a hit is lower than the values given in Table \ref{Hng}.

\begin{table}
\begin{tabular}{c*{6}{c}}
\hline Length   & \multicolumn{6}{c}{$\langle$G4\%$\rangle$} \\
        	& p             & $\alpha$      & Li
        	& C         	& N         	& O \\
\hline
        2   &3.8$^{+2.7}_{-2.7}$    &5.4$^{+1.6}_{-1.6}$    &5.6$^{+2.2}_{-2.2}$
            &7.5$^{+2.6}_{-2.6}$    &6.2$^{+1.4}_{-1.4}$    &7.6$^{+1.8}_{-1.8}$    \\
        3   &0.6$^{+1.2}_{-0.6}$    &0.9$^{+2.1}_{-0.9}$    &0.7$^{+0.9}_{-0.7}$
            &1.0$^{+1.3}_{-1.0}$    &1.8$^{+1.1}_{-1.1}$    &1.0$^{+0.8}_{-0.8}$    \\
        4   & -             	    & -         	    & -
            &0.3$^{+0.5}_{-0.3}$    &0.1$^{+0.3}_{-0.1}$    &0.1$^{+0.3}_{-0.1}$    \\
\hline
\end{tabular}
\caption[]{\small First column: number of pixel affected by the negative glitch. Other columns: the mean
percentage (5 simulations) of GCRs affecting a fixed pixel length, in the shield+secondary events
configuration, for each GCRs type, and for the high stressed PhC. Similar values are found for low stressed
PhC, and at solar maximum.}\label{Hng}
\end{table}

\subsection{Secondary Events}\label{secondary}
The formation of secondary events depends strictly on the physics we considered. In case of proton GCRs
we obtained a lot of secondary events, from photons and electrons to a large sample of particles
(i.e. $\pi^+$) to some nuclei (i.e. $^{27}$Al). For the light nuclei, secondary events
are basically photons or electrons. Ion inelastic scattering may also activate and generate pions,
protons and electrons and smaller nuclei, but these events are much rarer.

\begin{figure}
\centerline{\includegraphics[height=8cm, width=8cm]{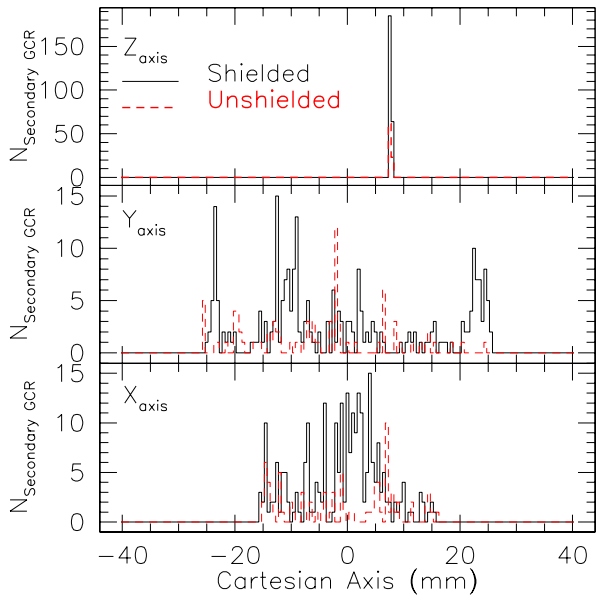}}
\caption[]{\small Cartesian projection of secondary events generation space for the high stressed PhC
(from 5000 protons seed) for shielded (straight line) and unshielded (dashed line) configurations. Upper
panel; the projection along the Z axis, mid and lower panels projection along the Y and
X-axis.}\label{xyzsec}
\end{figure}

Figure \ref{xyzsec} reports the number of impacts on the detector in a shielded and unshielded
configuration (high stressed PhC). The shield does not act as the major source of secondary events
and does not largely affect the pixels. Ge pixels are well embedded within other detector
components masking the shield, the origin of the secondary. 

As expected, the $n_{\rm sec}/n_{\rm prim}$ ratio is high because of the large mass of the detector. The
ratio is almost identical for high and low stressed PhC, since the two arrays are very similar.

\section{Proton irradiation tests}\label{IPN}
At the Light Ion Facility of the Centre de Recherches du Cyclotron of the University Catholique de Louvain
La Neuve, Belgium, some tests were performed, intended to determine the glitch event rate on a single
photoconductor module. The 2004 tests could not be reproduced because of sudden unexpected responsivity
changes which made impossible to follow the detector behaviour. In addition the external environment was
not clearly defined and simulations could not usefully compared to measurements. Tests were repeated in
April 2005 and the responsivity jumps were cured by heating the detectors (see Katterloher et al. 2005).
Here we report on these latter tests.

\subsection{Test Specimen}
\begin{figure}
\centerline{\includegraphics[height=12cm, width=8cm]{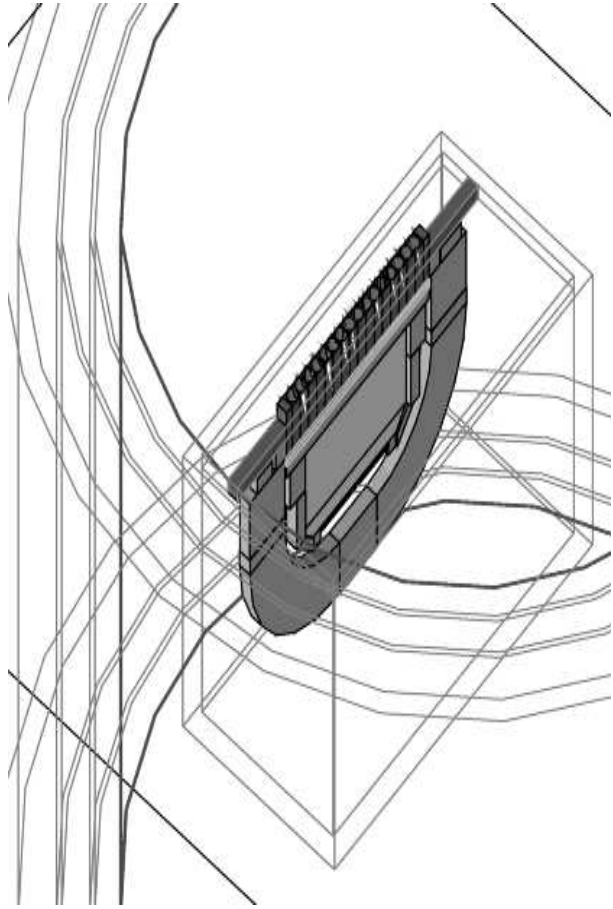}}
\caption[]{\small The dewar, and a single stressed module, in the Louvain cyclotron configuration. In
gray the three Al cylinders and the box around the module, in red the Cu cylinder foil.}\label{dewar}
\end{figure}

The test specimen foreseen for the proposed investigations under proton irradiation is the detector
module FM\#12 with a CRE of the Qualification Model type, mounted in the centre of the module. The
detector is in the high stress configuration. The module stood in a liquid He dewar operating at a
temperature of (1.85$\pm$0.05) K.\\
The dewar is made of three concentric cylinders of Al (called hereafter Al1, Al2 and Al3) and one made of
Cu. Al1 is 6 mm thick.
Al2 and Al3 are 1.5 mm thick. 
The Cu cylinder is 0.5 mm thick. 
The module is placed inside a box of Al 4 mm thick ({\it Al case}) so that half the fore-optics stands
outside of it (see Fig. \ref{dewar}). The module is not placed at the centre of the cylinders.

The proton beam, before reaching the Ge:Ga crystals elements, penetrates two layers, one made of steel, 60
$\mu$m thick, and one made of lead 0.12 mm thick. There is also a couple of steel collimators, that we
modeled as a box, with a hole in its centre as large as the beam.

\begin{figure}
\centerline{\includegraphics[height=12cm, width=12cm]{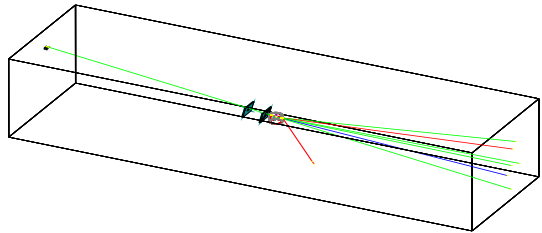}}
\caption[]{\small The UCL-CRC experimental hall. The detail of the DUT (here in the center of the plot)
is in Figure \ref{dewar}.}\label{specimen}
\end{figure}

During the test a primary circular (10 cm $\emptyset$) proton beam of 70 MeV was used, with a beam length
of 5.12 m between the diffusion foil and the Device Under Test (DUT). The beam reaches the dewar (and the
module) under an angle of 10$^{\circ}$ (see Figure \ref{specimen}). The DUT is placed into air. Due to air
leakage, we set air cuts as large as we can, in order not to have a massive secondary event generation,
that is 2.5 m.

\subsection{The G4 Simulations}
\begin{figure}
\centerline{\includegraphics[height=12cm, width=12cm]{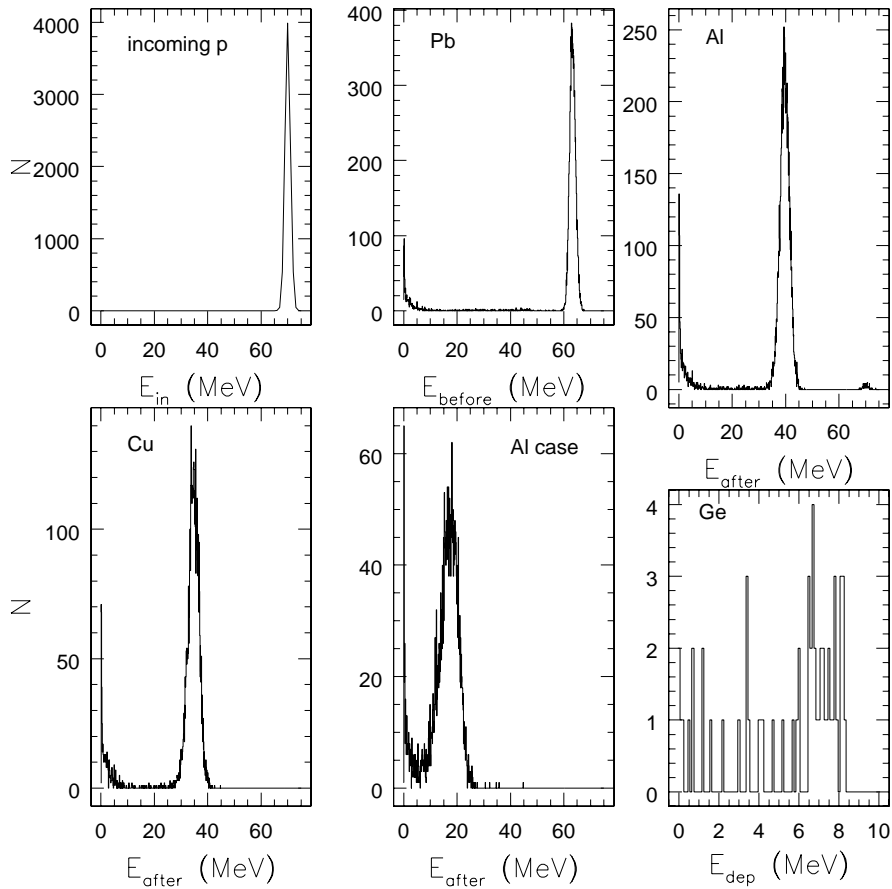}}
\caption[]{\small Degradation of the proton energy across the dewar layers: from the top-left panel, the
energy of the incoming proton flux, the flux emerging from the polystyrene foil, after the Aluminum
shields, after the Cu case and the final detector case. The last panel refers to the energy deposition
on the Ge:Ga material.}\label{estop05}
\end{figure}

We first ran 5 simulations of 10000 protons each, in order to have an idea of what occurs to the beam once
it crosses all the layers. We assume that the beam had a Gaussian error of 1 MeV. Before the first Al
cylinder layer we had an energy of $\approx$ 58.09 MeV, which disagrees with data by J. Cabrera, 63.74 MeV
\cite{Katterloher05}. This is due to the new G4 version we used (6.2.02\footnote{Furthermore the 6.7.01
version gave a very rare crash, probably due to a bug in the Hadronic Physics for the High Precision
neutrons: with this new version we switched off such a physics.}, gave $\approx$63.70 MeV). Our modelled
dewar is reliable and represents closely the dewar geometry. We do not have reasons to believe the Geant4
fails in this case and we set the energy hitting the module at 15.22$\pm$1.95 MeV (bottom right panel of
Figure \ref{estop05}). In Figure \ref{estop05} we plot the degradation of the proton energy along the path
inside the dewar.\\
What we would like to stress here is that 10000 protons originate roughly 22300 secondary events. This is
mainly due to the air leakage. After the Al box, $\approx$610 protons survive. But, after the Al box,
there are also $\approx$330 secondary events. The spectral range and type of these secondary events is
strictly dependent of the reliability of the G4 models.

In Table \ref{p-rate05} we give the relevant information on the UCL-CRC experiment as performed at the
Louvain cyclotron \cite{Katterloher05}.

\begin{table}
\begin{tabular}{rrrrrrrrr}
\hline
Exp.    & BIAS  & t$_{ramps}$   & C & N$_{ramps}$    & N$_{rep}$     & Flux         \\
        & (mV)  & (ms)      & (pF)  &        &       & (cm$^{-2}$s$^{-1}$)  \\
\hline
 \#L1   &50 &250    &1.09   &1024   &35 & 10    \\
  	&50 &250    &1.09   &1300   & 3 & 10    \\
    	&30 &250    &0.23   &1024   &23 & 10    \\
        &20 &250    &0.23   &1024   & 4 & 10    \\
        &40 &250    &0.23   &1024   & 1 & 10    \\
        &50 &250    &0.23   &1024   & 1 & 10    \\
        &70 &250    &0.43   &1024   & 1 & 10    \\
        &30 &250    &0.23   & 512   & 2 & 10    \\
\hline
\#H1    &30 &250    &0.23   &1024   &18 &400    \\
\hline
\end{tabular}
\caption[]{\small Summary of proton seed and photo-conductor module setup during UCL-CRC experiments.
Column \# 1 reports the identifier of the chosen setup, \# 2 the value of the bias voltage of the FEE
circuit, \# 3 the duration of each integration ramp, \# 4 the integrator capacitor, \# 5 the number of
ramps, \# 6 the number of repetition with the same measurement setup, \# 7 the proton
flux.}\label{p-rate05}
\end{table}

\subsection{The Geant4 results}\label{cave}
\begin{figure}
\centerline{\includegraphics[height=8cm, width=8cm]{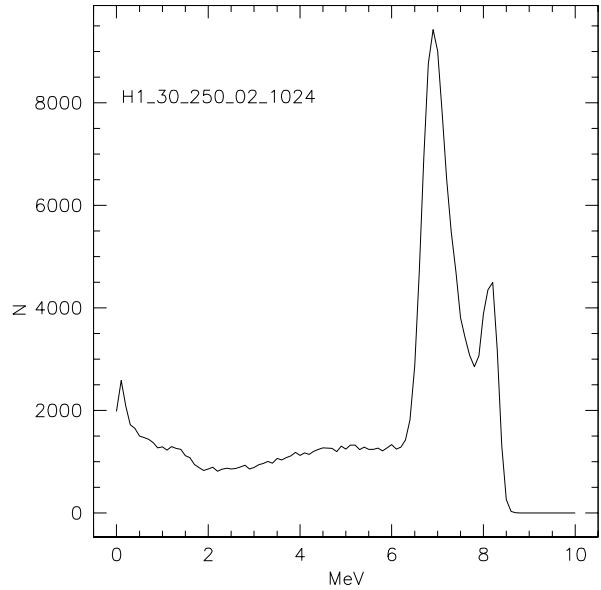}}
\caption[]{\small Energy deposition for the H1 test.}\label{H1}
\end{figure}

We ran the same experiments under the G4 tool. Results are summarized in Table \ref{G4out05}. In Fig.
\ref{H1} we plot the energy deposition onto the 16 pixels in the H1 test.\\
Here we found 3 peaks. The major one, is clearly due the primary proton beaming. The one at low
energies, is due to the secondary events generated along the tracking of the protons along all the
components we described before. The peak at higher energy could be due to the fact that the pixel are
inside a non uniform cavity: due to the inclination of the beam there is a part of it that cross the
cavity in its thinner part, that is: pixel are hit by more energetic protons.

\begin{table}
\begin{tabular}{rrrrrrrr}
\hline
Exp.    & N$_{rep}$ & Geant4$_{rep}$     & Flux         & t$_{rad}$ & N$_p$     &$\langle events\rangle$& Rate  \\
        &       &        & (cm$^{-2}$s$^{-1}$)  & (s)&          &   & (s$^{-1}$cm$^{-2}$)   \\
\hline
\#L1    &35 &35 & 10    &256    & 200960    & 251.83    &  3.53 \\
   	& 3 & 6 & 10    &325    & 255125    & 328.00    &  3.63 \\
    	&23 &23 & 10    &256    & 200960    & 252.87    &  3.55 \\
    	& 4 & 8 & 10    &256    & 200960    & 255.38    &  3.58 \\
    	& 1 & 5 & 10    &256    & 200960    & 267.20    &  3.75 \\
    	& 1 & 5 & 10    &256    & 200960    & 250.60    &  3.52 \\
    	& 1 & 5 & 10    &256    & 200960    & 253.20    &  3.55 \\
    	& 2 &10 & 10    &128    & 100480    & 129.10    &  3.62 \\
\hline
\#H1    &18 &18 &400    &256    &8038400    &9948.06    &139.58 \\
\hline
\end{tabular}
\caption[]{The UCL-CRC experiment outputs with Geant4.}\label{G4out05}
\end{table}

\subsection{Comparison with measurements}
A preliminary analysis of these data was made by Groenewegen \& Royer (2005). Here we report the results in
energy; the detector output signal in Volts can be retrieved from equation \ref{EdV}. In the report by
Groenewegen \& Royer (2005) the conversion value between MeV and Volt is taken to be 1.34.

\subsubsection{Events and Rates}
The number of events observed in files L26-L29 were 755, corresponding to a rate of 3.1 $s^{-1}cm^{-2}$. If
corrected for the efficiency of the detection algorithm it becomes 3.9 $s^{-1}cm^{-2}$. For such a dataset,
G4 finds 853 events and a rate of 3.53. The number of events observed in files H3-H6 is 29008, G4 has 29430
events. Within the uncertainties these values are similar. The fundamental physical processes occurring in
this experiment are therefore reproduced.

\subsubsection{Deposited Energy}
\begin{figure}
\centerline{\includegraphics[height=8cm, width=8cm]{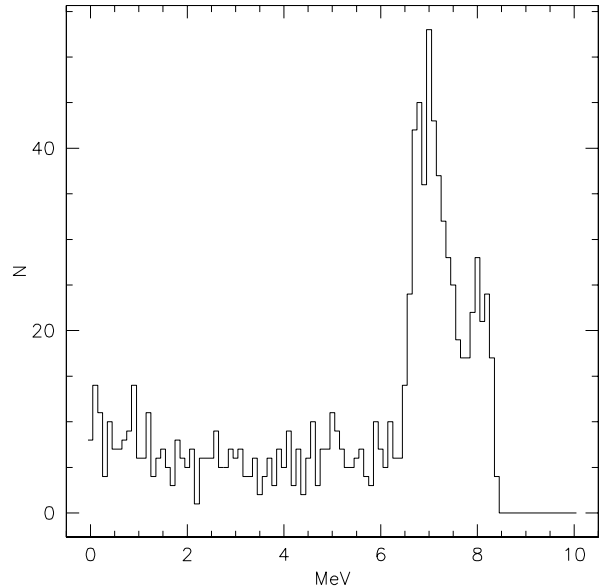} }
\caption[]{The energy distribution of glitches for files L26-L29, with Geant4.}\label{mv}
\end{figure}

In Figures \ref{H1} and \ref{mv} we plot the distribution of the predicted deposited energy for the files
L26, L27, L28 and L29 and those at large proton flux H1 in MeV. The energy distribution of the glitches,
i.e. that of the number of hits reaching the detector module, shows that many glitches occur at low energy
and the distribution is asymmetric with a peak at $\approx$7 MeV. The distributions are bimodal and have a
second peak around 8 MeV (see \S \ref{cave}) The energy glitch distribution is, as expected, the same in
both experiments (H1 and L26-L29) with low proton and high proton fluxes (Figures \ref{H1} and \ref{mv}).
What is changing is only the number of hits which in the second case is hundreds times lower. In fact, the
proton energy is the same, in both configurations, only the beam intensity is changing.

\begin{figure}
\centerline{\includegraphics[height=8cm, width=8cm]{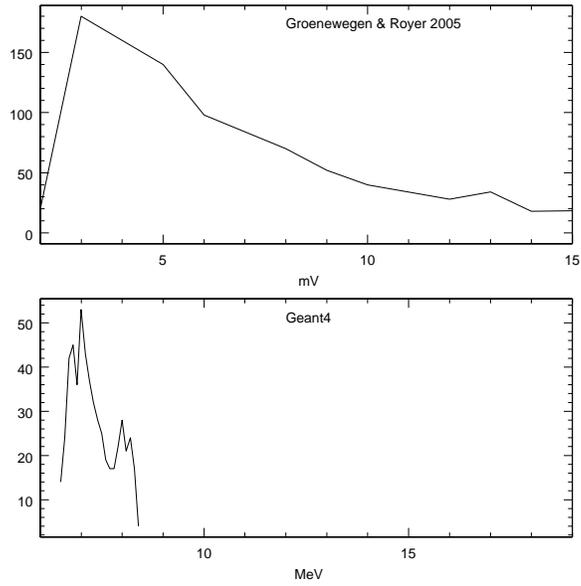} }
\caption[]{Upper panel: the observed distribution of glitches once a deglitching algorithm has been
applied. Lower panel: the G4 output of the same measuring setup and the same deglitching
algorithm.}\label{comparison}
\end{figure}

Figure \ref{comparison} shows, in the upper panel, the measured glitch distribution in Volt, and in the
lower panel the high energy tail of the glitch distribution shown in Figure \ref{mv}. We apply here the
same deglitching algorithm and get rid of the low energy tail of the hits. This figure, i.e. the glitch
detection, depends strictly on the algorithm used to de-glitch the data (Table 9 of Groenewegen \& Royer
2005). If, for instance, we applied a cutoff value around $\approx$6 MeV the distribution of glitches would
be more similar to the observed one. This latter has a peak at low energy, with a rather long tail towards
high energies (Groenewegen \& Royer 2005). The simulated distribution gave first the tail, then the two
peaks and is much narrower (cfr. \S\ref{cave}).

To compare the two distributions a conversion factor to the x-axis values (either to MeV or to Volt) must
be applied. This can be accomplished using eq.~(\ref{mv}). The resulting value is, in our case, $\Delta E
(MeV) = 6.71 \Delta V(V)$. This value was computed from the measurements taking into account the
responsivity changes under irradiating conditions (Lothar Barl, private communication). The two
distributions however differ substantially, the simulated one being much narrower. The difference could be
due to a large number of secondary events produced during the tests by some unknown components we are not
aware of. This component should have a large energy and it is unlikely that it occurs but this hypothesis
cannot be discarded. Another possible explanation is that the conversion factor from MeV to Volt is a
function of the energy. Although plausible, at present we cannot prove this possibility.

\subsubsection{Boundary Effect}
\begin{figure}
\centerline{\includegraphics[height=8cm, width=8cm]{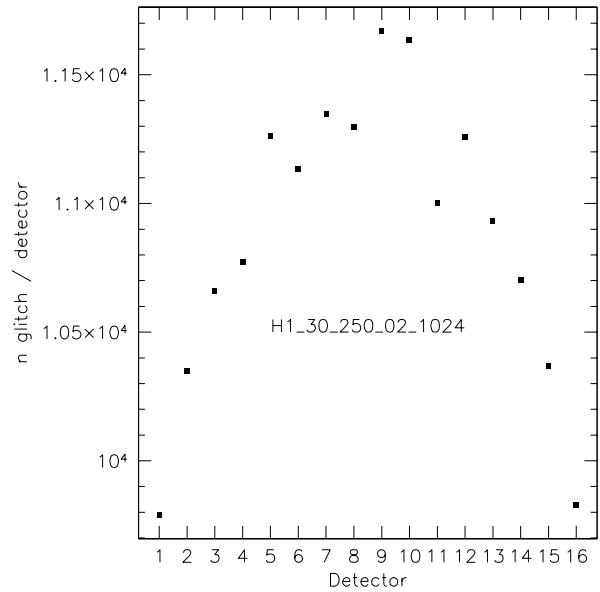}}
\caption[]{The boundary effect. Number of G4 simulated glitches per detector (pixel) in the H1
setup.}\label{bound05}
\end{figure}

Figure \ref{bound05} shows the number of detected glitches per detector: a clear {\it boundary effect}
(i.e. external pixels of the module were under-hit with respect to the central ones) is seen, and is due to
a differential incident proton flux with respect to the detector position (Figure 12 in Katterloher et al.
2005). We find a difference of the hit numbers of 20\% while the measured beam intensity difference between
the central pixel and the outer ones is 10\%. This is a geometrical effect and does not correspond to a
different behaviour of the single chip.

\section{Discussion}\label{discussion}
\subsection{The photoconductors in the space environment}
Our simulations of the effects of Galactic Cosmic Ray impacts on the photoconductor arrays show that both
the satellite plus the cryostat (the shield) and the detector act as source of secondary events, affecting
the detector performance. Secondary event rates originated within the detector and from the shield are of
comparable intensity. The impacts deposit energy on each photoconductor pixel but do not affect the
behaviour of nearby pixels. These latter are hit with a probability always lower than 7\%.\\ The energy
deposited produces a spike which can be hundreds times larger than the noise. The present simulations are
not able to follow the temporal behaviour of each pixel and cannot be used to determine the shape of the
output signal as a function of time.

\subsection{The photoconductors in the lab test}
We have simulated the experiment carried out at UCL-CRC and compared the simulation outputs to the
measurements reported in Groenewegen \& Royer (2005). We find similar rates and events (see Tables
\ref{MGvsG4}) on each pixel. The simulated energy distribution differs substantially from that of the
measurements.

We tried to ascribe these differences to the following causes:
\begin{enumerate}
\item As far as the event number and rate is concerned, it may be possible that the chosen cuts are rather
large. But due to the complexity of the experimental hall it is hard to tune them optimally. We would need
information on the intermediate passage of the beam through matter. A simple correction could be a slight
increase of the Ge cuts. A clear benchmark would be the comparison between the energy distributions.
\item The presence of a primary and a secondary peak seems to be the only feature common between
experimental and simulated data. What it is not clear is why, due to the crossing through matter, the
Gaussian beam is ({\it should be}) transformed as in Figure \ref{estop05}, that is first a tail, then a
peak, whereas experimental data favour the tail next to the secondary peak. Either Ge has some physical
properties which are not included in the G4 Physics List, or the deglitching algorithm used is not
correct.
\end{enumerate}

\begin{table}
\begin{tabular}{rrr}
\hline
	    & \cite{MGb}    & Geant4    \\
Files       & Events    & Events\\
\hline
L26-L29     &   755 &  853  \\
H3-H6       & 29008 &29430  \\
\hline
\end{tabular}
\caption[]{Measured versus predicted values.}\label{MGvsG4}
\end{table}

\begin{acknowledgements}
We warmly thank M. Asai, G. Cosmo, A. De Angelis, F. Lei, V. Ivantchenko, G. Santin and J.P. Wellish,
who helped us in this effort and elucidating us the tricks of the Geant4 code.
We thank Lothar Barl for his useful information about the photoconductor detector.

\end{acknowledgements}

\end{article}
\end{document}